# Controlled Growth of ZnO Nanowire, Nanowall, and Hybrid Nanostructures on Graphene for Piezoelectric Nanogenerators

*Brijesh Kumar,[†] Keun Young Lee,[†] Hyun-Kyu Park,[†] Seung Jin Chae,[‡] Young Hee Lee,[‡] &*

*Sang-Woo Kim[\*,†,§]*

School of Advanced Materials Science and Engineering, Sungkyunkwan University, Suwon 440-746, Republic of Korea, Department of Physics, Department of Energy Science, SKKU-Samsung Graphene Center, Sungkyunkwan University, Suwon 440-746, Republic of Korea, SKKU Advanced Institute of Nanotechnology (SAINT), Center for Human Interface Nanotechnology (HINT), SKKU-Samsung Graphene Center, Sungkyunkwan University, Suwon 440-746, Republic of Korea

[†] School of Advanced Materials Science and Engineering, Sungkyunkwan University.

[‡] Department of Physics, Department of Energy Science, SKKU-Samsung Graphene Center, Sungkyunkwan University.

[§] SKKU Advanced Institute of Nanotechnology (SAINT), Center for Human Interface Nanotechnology (HINT), SKKU-Samsung Graphene Center, Sungkyunkwan University.

[\*] Corresponding author E-mail: (S.-W. Kim) kimsw1@skku.edu

**Abstract**

**Precise control of morphologies of one-dimensional (1D) or 2D nanostructures during growth has not been easily accessible, usually degrading the device performance and therefore limiting applications to various advanced nanoscale electronics and optoelectronics. Graphene could be a platform to serve as a substrate for both morphology control and direct use of electrodes due to its ideal monolayer flatness with π electrons. Here, we report that by using graphene directly as a substrate, vertically well-aligned ZnO nanowires and nanowalls were obtained systematically by controlling Au catalyst thickness and growth time, without invoking significant thermal damage on the graphene layer during thermal chemical vapor deposition of ZnO at high temperature of about 900 $^o$C. We further demonstrate a piezoelectric nanogenerator that was fabricated from the vertically aligned nanowire-nanowall ZnO hybrid/graphene structure generates a new type of direct current.**

Piezoelectricity, conversion of mechanical energy to electrical signals, is one of the most versatile phenomena to harvest energy to power small-scale electronic devices from environment. ZnO has several key advantages in this area, being a biologically safe piezoelectric semiconductor with a wide range of nanostructures such as nanowires, nanorods, nanotubes, etc.[1-3] The existence of various one-dimensional (1D) and 2D forms of ZnO opened opportunities for applications not only to energy harvesting but also to optoelectronics and photovoltaics.[4,5] Nevertheless, the applications have been limited due to the poor control of self-assembled nanostructures that often lead to degrade the device performance and make them inaccessible to many application areas.

Graphene is a two-dimensional (2D) system composed of carbon atoms arranged in a hexagonal, honeycomb lattice with a unique electronic structure; its excellent optical transparency, mechanical flexibility, thermal stability, and chemical inertness make it an ideal material for the realization of transparent electrodes, solar cells, photodetectors, nanogenerators, and light-emitting diodes.[6-10] Due to its linear energy-momentum dispersion relation, graphene has unique electronic structures of relativistic massless Dirac particles, high electron mobility approaching 200,000 $cm^2V^{-1}s^{-1}$, and excellent optical transmittance of ~ 97%, while maintaining high mechanical strength and thermal conductivity similar to carbon nanotubes.[6,11]

1D (or 2D) semiconducting nanostructure/graphene system can offer a unique opportunity to study the physics at interfaces between semiconducting nanostructures and graphene. For example, semiconductor-graphene junctions with high-quality crystal structure can be ideal to lead to effective carrier transport at the interface between two nanostructures with significantly reduced carrier scattering or traps, as well as novel device structures that have not been accessible in isolated graphene and semiconducting nanostructures. In this respect, the platform of 3D nanoscale building blocks via the integration of semiconducting nanostructures with 2D graphene layers is very promising for the realization of graphene-based functional nanodevices.

Another usefulness of graphene layer is that graphene layer can be directly used to grow nanostructures with high crystallinity and used directly as an electrode for devices. ZnO nanostructures in 1D or 2D have been grown previously on highly oriented pyrolytic graphite using catalyst-assisted chemical vapor deposition (CVD)[12] and on graphene sheets using metalorganic CVD (MOCVD) and a solution approach.[13,14] However, high-quality ZnO nanowires with high density, high aspect ratio, large-area uniformity, and perfect vertical alignment have not been obtained, let alone the issue for the damage of graphene layers.

The purpose of this work is two-fold: i) To grow directly high-quality ZnO nanostructures on graphene layers in a controllable way without invoking damage on graphene layers and ii) to apply this structure to a graphene/ZnO-based direct current (DC) nanogenerator. The growth mode of ZnO on CVD-grown graphene layers was controlled precisely by varying Au layer thickness and growth time using thermal CVD method. Our study shows that with increasing Au thickness, the ZnO morphology gradually changed from a pure nanowire to a pure nanowall structure, through a hybridized state composed of an interconnected nanostructure of nanowires and nanowalls, where the nanowires were perfectly aligned vertically on (002) graphene surface at high temperature (900 $^{\circ}$C). We found that the piezoelectric nanogenerator based on the nanowires-nanowalls hybrid generated a DC type voltage. A DC type ZnO nanogenerator has been demonstrated before by lateral mechanical compression on bent nanowires,[15] while the effect here is achieved through the specific electron dynamics in the nanowire-nanowall hybrid.

The morphologies of the ZnO nanostructures were examined by field-emission scanning electron microscopy (FE-SEM). The structural properties of the samples were determined by X-ray diffraction (XRD) measurements and high resolution transmission electron microscopy (HR-TEM). The ZnO/graphene heterojunctions were further investigated using micro energy-dispersive X-ray spectroscopy (µ-EDS). The Au top electrode of the piezoelectric generator was deposited on a plastic substrate by a thermal evaporator and then pressed onto the ZnO structure. A Keithley 6485 picoammeter and 2182A voltmeter were used to measure the low-noise output current and voltage generated from the piezoelectric nanogenerators.

The tilted-view FE-SEM images in Figure 1a-c show morphologies of the vertically aligned ZnO nanowires, nanowire-nanowall hybrid, and nanowall structures on the grapahene/$Al_2O_3$ substrates that were grown with various Au layer thicknesses of 0.5 nm, 1 nm, and 2 nm, respectively, for 60 min at 900 °C. From the cross-sectional FE-SEM images shown in Figure 1d-f, the average diameter and height of the nanowires were estimated to be 90 nm and 3.0 μm, respectively, and those of the nanowalls were 200 nm and 2.4 μm, respectively. The observed diameter of the nanowires was approximately 90 nm and the thickness of the nanowalls was about 200 nm in the nanowire-nanowall hybrid structure. In the hybrid structure, nanowires were grown directly from the nodes of the nanowalls.[12] The ZnO nanowire, hybrid, and nanowall structures were grown by varying the growth time among 40, 60, and 120 min at 900 °C. The height of the nanowires increased from 1.5 μm to 5.0 μm with increasing growth time from 40 min to 120 min. On the graphene surface, Au–Au interaction is stronger than Au–graphene interaction.[16] Consequently, the Au layer is initially not continuous; as the thickness increases, a more continuous network of dense Au nanoparticles on the graphene surface resulting in the transformation of the nanowires into the nanowall structure during the ZnO growth.

The morphology of the ZnO nanostructures can be easily controlled by varying the Au layer thickness or the density of the Au nanoparticles on the graphene. As the Au layer thickness was increased from 0.5 nm to 2 nm, the density of the Au nanoparticles also increased. As a result, we successfully controlled the ZnO nanostructure morphologies (nanowire, nanowire-nanowall hybrid, and nanowall structures) by manipulating the Au particle density on the graphene substrates, as shown in Figure 1. The aspect ratio (length/diameter) and density of the vertically grown nanowires were 34 and $5 \times 10^9$ cm$^{-2}$, respectively. These values are comparable to those of nanowires grown on single-crystalline Si and $Al_2O_3$ substrates.[17,18] The aspect ratio of nanowires grown for 120 min increased to 55 with a surface density of $5 \times 10^9$ cm$^{-2}$, which is very large compared to nanowires grown on graphene by other methods, including MOCVD.[13]

The diffraction peaks (002) and (004) corresponding to the hexagonal (00l) ZnO phase in the XRD patterns indicate that the nanowire, hybrid, and nanowall structures were grown vertically and highly *c*-axis oriented (see Figure S1 of the supporting information). The cross section HR-TEM and μ-EDS results clarify the Au positions at the graphene-ZnO interface and reveal that Au nanoparticles decorated on graphene did not act as a catalyst in the growth process but acted as nucleation sites for the growth of ZnO nanostructures (see Figure S2 of the supporting information). This is due to the small sized Au nanoparticles and the perfect in-plane crystalline relation of Au (111) with graphene (002)[19] that forces the Au nanoparticles to bind with the graphene layer during the ramping prior to ZnO growth initiation.

The vertical *c*-axis growth of the ZnO nanowire, nanowire-nanowall hybrid, and nanowall structures was observed as a result of ZnO (002) nanostructures grown with the graphene (002) plane, as illustrated in the cross-sectional HR-TEM image of the nanowire-nanowall hybrid structure in Figure 2. The lattice spacings of 0.52 nm and 0.38 nm, calculated from Figure 2, are consistent with the values of ZnO (002) and graphene (002), respectively.[9] In Figure 2c, the fast Fourier transform (FFT) of region (I) clearly reveals that graphene (002) and ZnO (002) planes are oriented in the same direction, while the *c*-Al$_2$O$_3$ (002) plane is tilted from the (002) planes of graphene and ZnO. This result indicates that graphene can be a novel platform to serve as a buffer layer for realization of ZnO nanostructures that are perfectly aligned perpendicular from any kinds of substrates.

The above results provide detailed features of the growth process of ZnO nanowire, nanowall, and hybrid nanowire-nanowall structures allowing for the elucidation of the growth mechanism of graphene-based ZnO nanostructures. The growth mechanism of the nanowires is shown in Figure 3a. At First, Zn vapor is generated by the thermal carbon reduction of ZnO in the source region at high temperature. The metallic Zn vapor is then transferred to the Au-decorated graphene. A portion of the Zn vapor is adsorbed directly onto the Au nanoparticles on the graphene surface, and the rest of the Zn vapor produces droplets on the graphene. Zn droplets that migrate from the graphene surface to the nano-sized Au droplets act as energetically favorable sites for ZnO nanostructure growth. This is due to the higher sticking coefficient of ZnO with the liquid Au droplets, and the lower sticking coefficient with the solid graphene surface. As a result, Zn droplets preferably migrate from the graphene surface to the Au droplets (step I).

Secondly, due to the large quantity of Zn that has migrated from the graphene surface and the continuous supply of Zn prior to Zn and Au becoming soluble, further deposition produces the outer layers that encapsulate the Zn-adsorbed Au nanoparticles. The perfect epitaxial relation of Au (111) with graphene (002) forces the Au nanoparticles to bind with the graphene layer (step II).[19] From the measurement of the in-plane diffusion of Au on graphene at 600 °C, Gan *et al.*[20] obtained a diffusion coefficient, $D$, between $6 \times 10^{-22}$ and $2 \times 10^{-21}$ m$^2$s$^{-1}$. The diffusion coefficient is related to the activation energy, $E_a$, required for atoms to jump by

$$D \propto \exp\left(-\frac{E_a}{K_b T}\right) \qquad (1)$$

where $T$ is temperature and $K_b$ Boltzmann's constant. In our experiments, the growth temperature of 900 °C is so high for Au atoms to heavily diffuse through the graphene surface, resulting in the formation of strong covalent bonds between Au and carbon atoms within the graphene network.

In the next step, the O component supplied from the residual air dissolves into the Zn liquid droplets and as a result, growth of the ZnO nanostructures from the energetically favorable Au sites occurs (step III). The continuous supply of Zn and O vapors saturates the outer layers, followed by the precipitation of the ZnO nanowires. Step IV shows how the ZnO nanowires grow from the outer layers through the self-catalytic vapor-liquid-solid (VLS) process. The outer diameter of the Zn droplets encapsulating the Au nanoparticles restricts the diameter of the ZnO nanowires within the nanometer range. The maximum number and diameter of ZnO nanowires depends on the density and diameter of the Au droplets. The Zn droplets encapsulating the Au act as a self catalyst in the VLS growth. The length of the ZnO nanowires increases with growth time.

ZnO nanowires grow if the density of Au nanoparticles is not too high and Au nanoparticles are well separated from each other. If the distance between the Au droplets that act as nucleation sites on the graphene is very short (less than the diameters of the nanowires) then the Zn droplets that encapsulate the Au and act as a self catalyst will quickly merge together and form a continuous network of Zn droplets. As a result, nanowalls grow through the same self-catalytic VLS process depicted in the schematic diagram shown in Figure 3b.

Previous research has shown that self-generated clusters from the source can exhibit the diffusion phenomenon along or within the nanowires during growth at high temperature.[21] In the nanowire-nanowall hybrid structure, the nanowalls grow first as a result of dense Au nanoparticles located on the graphene surface, and then nanowires grow from the nodes of the nanowall networks. On the inclined nanowall surface toward the nodes, Zn droplets drift and aggregate at the nodes. Due to the high surface energy at the node, Zn droplets accumulate at the nodes for overall energy compensation. Therefore, ZnO nanowires begin to form from the nodes through the self-catalytic VLS process at the critical point. Nanowires also grow from the less dense Au nanoparticles on the graphene surface in the hybrid structure through the same self-catalytic process, which is well described in Figure 3a. Figure 3c shows a schematic diagram of the time-dependent ZnO nanowire-nanowall hybrid structure growth mechanism. This study suggests that Au plays an important role in the formation of these useful and novel ZnO nanowire, nanowall, and nanowire-nanowall hybrid structures on graphene.

Here, we demonstrated DC type piezoelectric nanogenerators as a potential application of this novel nanowire-nanowall hybrid structure. The generated output was measured from the nanogenerator fabricated with Au-coated polyethersulfone (PES) as a top electrode and the remaining graphene layer as a bottom electrode. The pushing force (0.5 kgf) was applied to the top of the nanogenerator in the direction perpendicular to the electrode. Figure 4a and b show the current density and voltage generated from the ZnO nanowire-nanowall hybrid nanogenerator, respectively. The output current density was approximately 500 nAcm$^{-2}$ and the output voltage was about 20 mV. Despite the 400 Ω sheet resistance of the graphene, which is much larger than the 60 Ω of the commercially available indium tin oxide,[9,22] the DC output current was successfully and clearly detected. The output current peaks were sharp and narrow. We believe that the output power can be further improved by using graphene with a lower sheet resistance and a ZnO hybrid structure with a larger height.

To verify that the measured signal was from nanogenerators rather than the measurement system, we performed ''switching-polarity'' tests, as shown in Figure 4. When the current and voltage meters were forward connected to the nanogenerator, positive pulses were recorded during the pushing. When the current and voltage meters were connected in reverse, the pulses were also reversed. The output current density and voltage for both connecting conditions were almost the same. The measured voltage of 20 mV from the nanogenerator is much lower than the calculated piezoelectric potential of ~ 0.63 eV. The calculated potential is higher than the Schottky barrier height of the Au-ZnO interface (approximately 0.40 eV[23]). The piezoelectric potential generated in the ZnO hybrid structure is sufficient to drive the piezoelectric induced electrons from the top Au electrode to the bottom graphene electrode. The piezoelectric potential was calculated using the relation $V_{max} = Fg_{33}L/A$,[1] where $F$ is the force applied on the top electrode (0.5 kgf), $g_{33}$ is the piezoelectric voltage coefficient of the ZnO nanowires (0.135 Vm/N,[24] neglecting the change of the piezoelectric voltage coefficient by nanostructuring of ZnO), $L$ is length of the nanowires in the hybrid structure used to fabricate the nanogenerator (3 μm), and A is contact area of the nanowires (3.14 mm$^2$).

The following explanation may account for this discrepancy. First of all, the contact resistance may have been very large as a result of the small contact area between the ZnO nanowires and electrode. Therefore, the voltage created by the piezoelectric effect is largely consumed at the contact due to its large contact resistance and only a small portion is received as the output.[25] Secondly, the free carriers in the nanowalls screened the piezoelectric potential generated in the nanowires, which could be another reason for the low measured voltage and current. Figure 5 shows a schematic illustration of an integrated nanogenerator with a Au top electrode and a piezoelectric ZnO nanowire-nanowall hybrid structure grown on a graphene/$Al_2O_3$ substrate along with its working mechanism.

The work function of the graphene electrode was about 4.53 eV and the electron affinity of ZnO is 4.35 eV.[26,27] Hence, a weak Schottky contact is formed at the interface between the graphene and ZnO nanostructures. As we apply the force in the vertical direction on the top electrode of the nanogenerator, the top surface of the nanowires reveals negative potential ($V_1^-$) and the bottom surface shows positive potential ($V_1^+$). Subsequently, the force-induced piezoelectric potential at the top surface drives the flow of the electrons from the top electrode to the bottom electrode, which consequently produces the measured current pulse during pushing.

In the nanowire-nanowall hybrid structure, some of the positive potential is screened by the free electrons present near the nanowalls. Thus, the magnitude of $V_2^+$ will be reduced as the electrons move from the top electrode and penetrate into the ZnO area contacting the graphene electrode by passing through the weak Schottky barrier. As a result, this weak positive potential is not sufficient to drive back the electrons accumulated near the bottom electrode toward the top electrode. Therefore, no current pulses are measured in releasing the force, resulting in DC type power generation from the nanowire-nanowall hybrid structure-based nanogenerator.

In summary, *c*-axis oriented novel ZnO nanowire, nanowire-nanowall hybrid, and nanowall structures with the graphene (002) plane were grown for use in functional nanoscale electronics and optoelectronics devices. It was found that Au plays an important role in the formation of the ZnO nanowire, nanowall, and nanowire-nanowall hybrid structures on graphene. A new growth mechanism of the ZnO nanowire, nanowall, and nanowire-nanowall hybrid structures on electrical conducting graphene electrodes was proposed in detail. Furthermore, we demonstrate that a DC output power can be generated by the piezoelectric ZnO/graphene nanogenerator due to the electron dynamic specific of the nanowire-nanowall hybrid structure.

**Supporting Information.** Growth of graphene sheets and ZnO nanostructures methods, XRD patterns of ZnO nanostructures, HR-TEM image and the µ-EDS spectrum of the nanowire-nanowall hybrid structure on a graphene/$Al_2O_3$ substrate for Au nanoparticle mapping. This material is available free of charge via the Internet at http://pubs.acs.org.

**Corresponding Author**

*Tel: +82-31-290-7352. Fax: +82-31-290-7381. E-mail: kimsw1@skku.edu

ACKNOWLEDGMENT

This research was supported by the International Research & Development Program of the National Research Foundation of Korea (NRF) funded by the Ministry of Education, Science and Technology (MEST) (2010-00297) and by Basic Science Research Program through the NRF funded by the MEST (2009-0077682 and 2010-0015035). One of the authors (Y.H.L.) acknowledges financial supports by the STAR-faculty program and WCU (World Class University) program through the NRF funded by the MEST (R31-2008-000-10029-0), and the IRDP of NRF (2010-00429) funded by the MEST in 2010 in Korea.

**REFERENCES**

(1) Cha, S. N.; Seo, J.-S.; Kim, S. M.; Kim, H. J.; Park, Y. J.; Kim, S.-W.; Kim, J. M. *Adv. Mater.* **2010**, *22*, 4726.

(2) Choi, M.-Y.; Choi, D.; Jin, M.-J.; Kim, I.; Kim, S.-H.; Choi, J.-Y.; Lee, S. Y.; Kim, J. M.; Kim, S.-W. *Adv. Mater.* **2009**, *21*, 2185.

(3) Zhang, B. P.; Binh, N. T.; Wakatsuki, K.; Segawa, Y.; Yamada, Y.; Usami, N.; Kawasaki, M.; Koinuma, H. *Appl. Phys. Lett.* **2004**, *84*, 4098.

(4) Chen, M.-T.; Lu, M.-P.; Wu, Y.-J.; Song, J.; Lee, C.-Y.; Lu, M.-Y.; Chang, Y.-C.; Chou, L.-J.; Wang, Z. L.; Chen, L.-J. *Nano Lett.* **2010**, *10*, 4387.

(5) Wei, Y.; Xu, C.; Xu, S.; Li, C.; Wu, W.; Wang, Z. L. *Nano Lett.* **2010**, *10*, 2092.

(6) Bae, S. K.; Kim, H. K.; Lee, Y.; Xu, X. F.; Park, J. S.; Zheng, Y.; Balakrishnan, J.; Lei, T.; Kim, H. R.; Song, Y. I.; Kim, Y. J.; Kim, K. S.; Özyilmaz, B.; Ahn, J.-H.; Hong, B. H.; Iijima, S. *Nat. Nanotechnol.* **2010**, *5*, 574.

(7) Wang, X.; Zhi, L.; Müllen, K. *Nano Lett.* **2008**, *8*, 323.

(8) Mueller, T.; Xia, F.; Avouris, P. *Nat. Photon.* **2010**, *4*, 297.

(9) Choi, D.; Choi, M.-Y.; Choi, W. M.; Shin, H.-J.; Park, H.-K.; Seo, J.-S.; Park, J.; Yoon, S.-M.; Chae, S. J.; Lee, Y. H.; Kim, S.-W.; Choi, J.-Y.; Lee, S. Y.; Kim, J. M. *Adv. Mater.* **2010**, *22*, 2187.

(10) Chung, K.; Lee, C.-H.; Yi, G.-C. *Science* **2010**, *330*, 655.

(11) Zhang, Y.; Tan, Y.-W.; Stormer, H. L.; Kim, P. *Nature* **2005**, *438*, 201.

(12) Ng, H. T.; Li, J.; Smith, M. K.; Nguyen, P.; Cassell, A.; Han, J.; Meyyappan, M. *Science* **2003**, *300*, 1249.

(13) Kim, Y.-J.; Lee, J.-H.; Yi, G.-C. *Appl. Phys. Lett.* **2009**, *95*, No. 213101.

(14) Lee, J. M.; Pyun, Y. B.; Yi, J.; Choung, J.; Park, W. I. *J. Phys. Chem. C* **2009**, *113*, 19134.

(15) Wang, X.; Song, J.; Liu, J.; Wang, Z. L. *Science* **2007**, *316*, 102.

(16) Zhang, Y.; Franklin, N. W.; Chen, R. J.; Dai, H. *Chem. Phys. Lett.* **2000**, *331*, 35.

(17) Park, W. I.; Yi, G.-C.; Kim, M.; Pennycook, S. J. *Adv. Mater.* **2002**, *14*, 1841.

(18) Yuan, G. D.; Zhang, W. J.; Jie, J. S.; Fan, X.; Zapien, J. A.; Leung, Y. H.; Luo, L. B.; Wang, P. F.; Lee, C. S.; Lee, S.T. *Nano Lett.* **2008**, *8*, 2591.

(19) Lee, J.; Lee, J.; Tanaka, T.; Mori, H. *Nanotechnology* **2009**, *20*, No. 475706.

(20) Gan, Y.; Sun, L.; Banhart, F. *Small* **2008**, *4*, 587.

(21) Shi, J.; Grutzik, S.; Wang, X. *ACS Nano* **2009**, *3*, 1594.

(22) Choi, D.; Choi, M.-Y.; Shin, H.-J.; Yoon, S.-M.; Seo, J.-S.; Choi, J.-Y.; Lee, S. Y.; Kim, J. M.; Kim, S.-W. *J. Phys. Chem. C* **2010**, *114*, 1379.

(23) Das, S. N.; Choi, J.-H.; Kar, J. P.; Moon, K.-J.; Lee, T. I.; Myoung, J.-M. *Appl. Phys. Lett.* **2010**, *96*, No. 092111.

(24) Gautschi, G. Piezoelectric Sensorics; Springer-Verlag, Berlin, Germany, 2002.

(25) Wang, Z. L. *Adv. Funct. Mater.* **2008**, *18*, 3553.

(26) Yan, Q.; Huang, B.; Yu, J.; Zheng, F.; Zang, J.; Wu, J.; Gu, B.-L.; Liu, F.; Duan, W. *Nano Lett.* **2007**, *7*, 1469.

(27) Liu, Y. L.; Liu, Y. C.; Yang, H.; Wang, W. B.; Ma, J. G.; Zhang, J. Y.; Lu, Y. M.; Shen D. Z.; Fan, X. W. *J. Phys. D: Appl. Phys.* **2003**, *36*, 2705.

**FIGURE CAPTIONS.**

**Figure 1.** Tilted-view FE-SEM images of ZnO (a) nanowires, (b) nanowire-nanowall hybrid, and (c) nanowall structures. Cross-sectional FE-SEM images of ZnO (d) nanowire, (e) nanowire-nanowall hybrid, and (f) nanowall structures on a graphene/$Al_2O_3$ substrate.

**Figure 2.** (a) Cross-sectional bright-field TEM image of the nanowire-nanowall hybrid structure. (b) Cross-sectional HR-TEM image of the graphene-ZnO interface in the nanowire-nanowall hybrid structure. (c) The FFT of region (I) shows that the graphene (002) and ZnO (002) planes are oriented in the same direction.

**Figure 3.** Schematic diagram of the growth mechanism of ZnO (a) nanowire, (b) nanowall, and (c) nanowire-nanowall hybrid structures on grapahene/$Al_2O_3$ substrates.

**Figure 4.** (a) Output current density generated from the nanogenerator fabricated with the nanowire-nanowall hybrid structure during the switching-polarity test. (b) Voltage generated from the nanogenerator fabricated with the nanowire-nanowall hybrid structure during the switching-polarity test.

**Figure 5.** Schematic illustration of an integrated nanogenerator with a Au top electrode (Au-coated PES substrate) and its working mechanism. (a) The as-received nanogenerator with no force application. (b) Electrons flow from the top electrode to the bottom side through the external circuit by the negative piezoelectric potential generated at the top side of the nanowire-nanowall hybrid structure under direct compression in the vertical direction. (c) Some of the positive potential is screened by the free electrons present near the nanowalls. The electrons moved from the top electrode penetrate into the ZnO area contacting the graphene electrode by passing through the weak Schottky barrier.

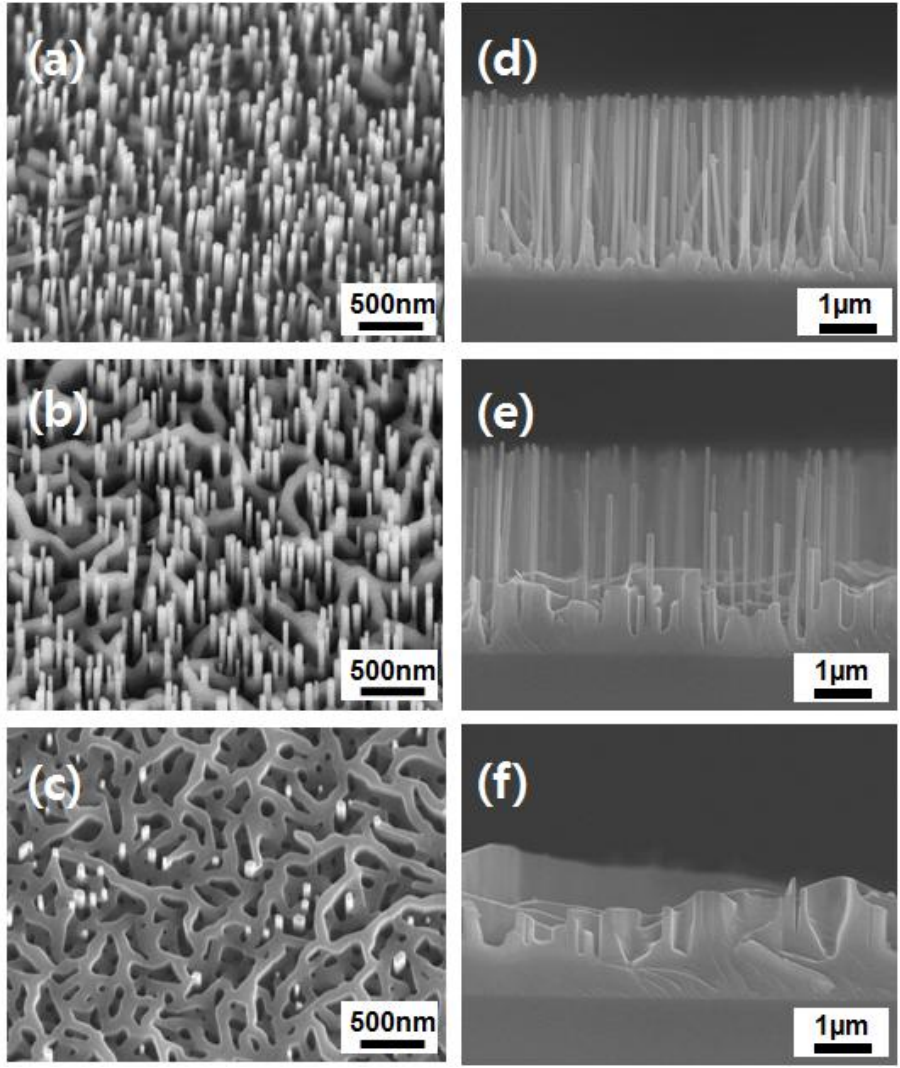

**Figure 1. Kumar et al.**

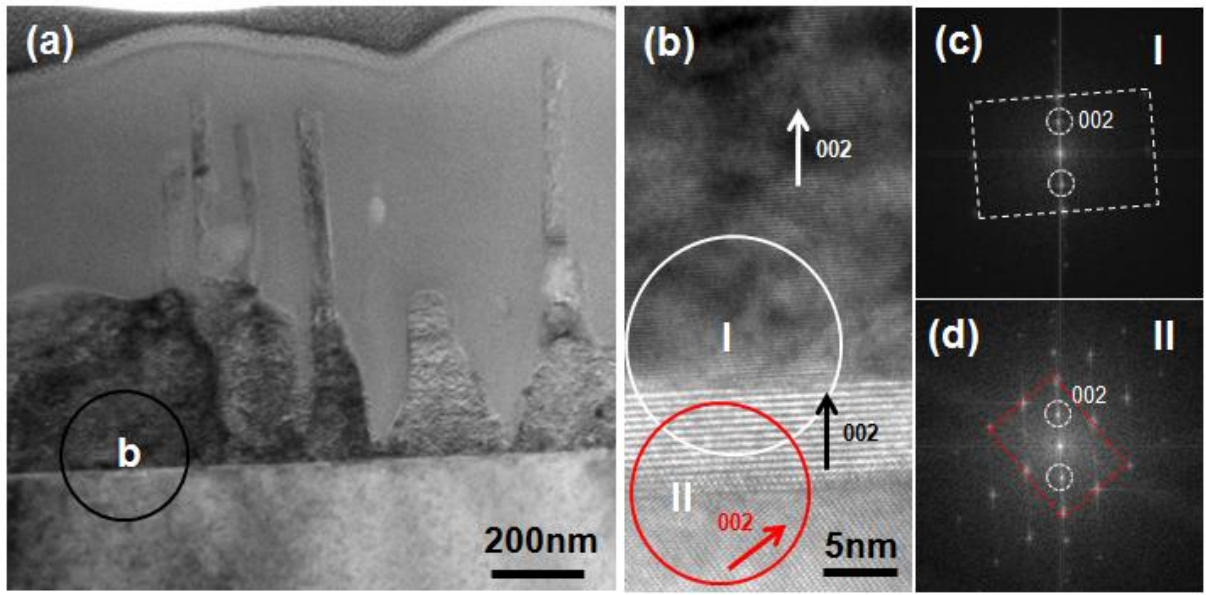

**Figure 2. Kumar et al.**

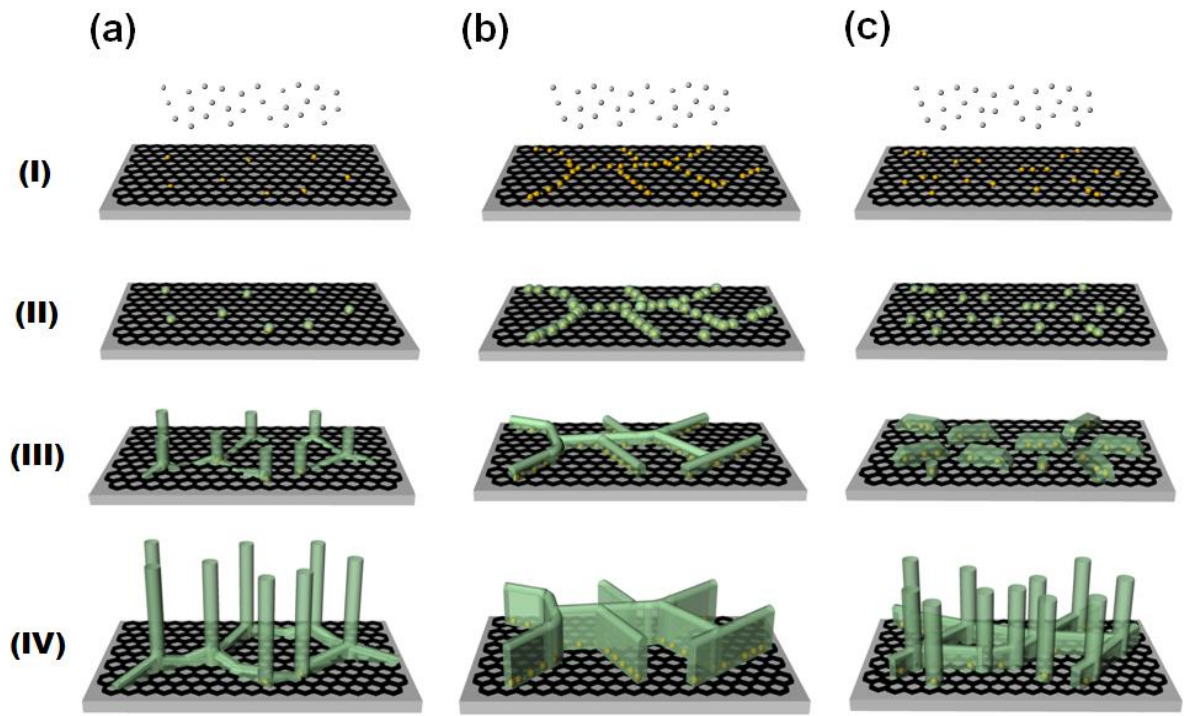

**Figure 3. Kumar et al.**

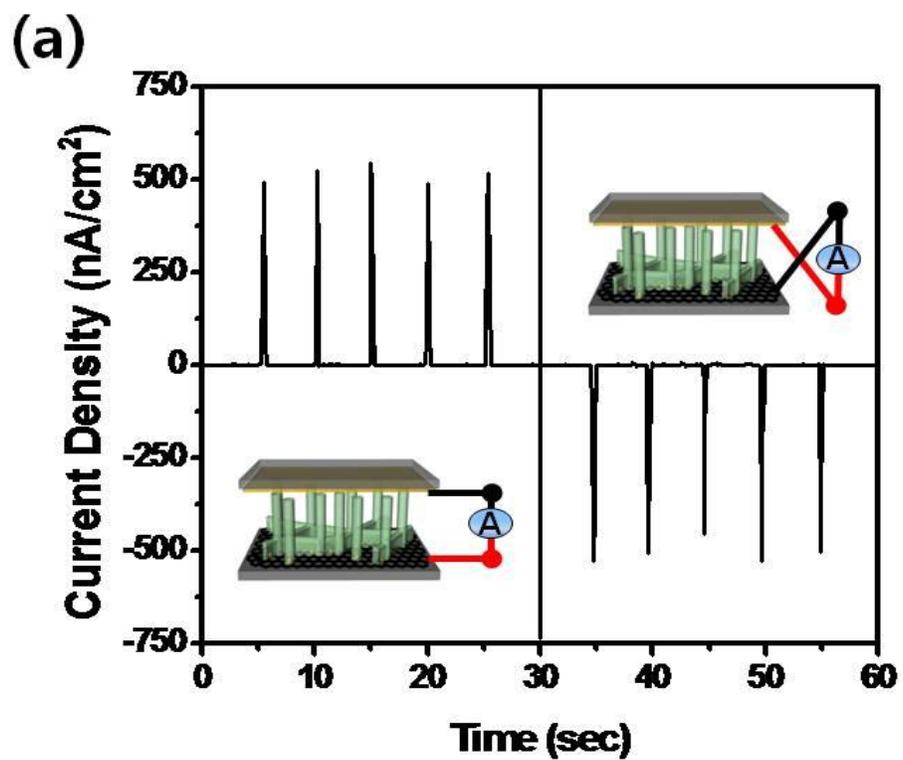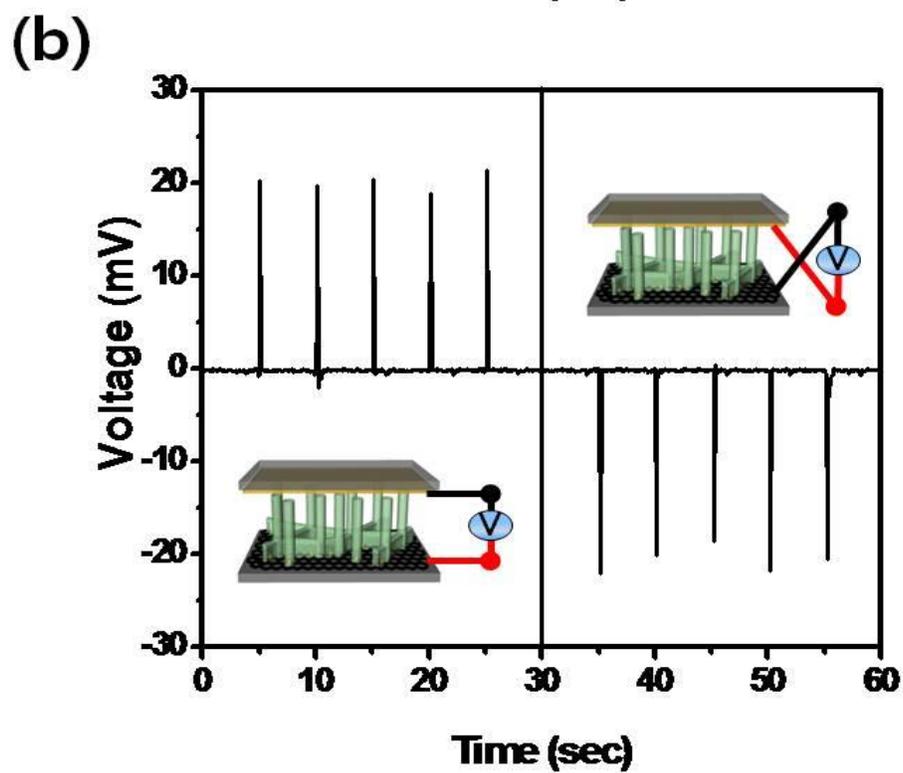

**Figure 4. Kumar et al.**

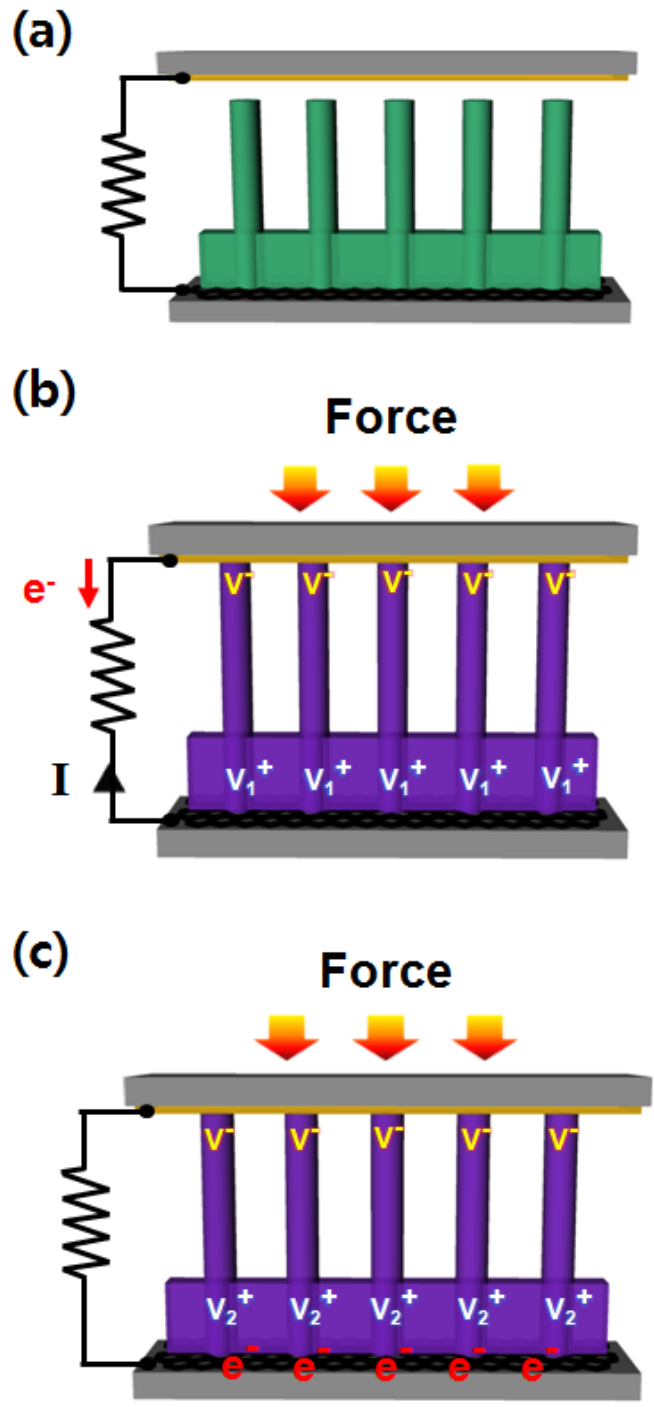

**Figure 5. Kumar et al.**

**Synopsis of table of contents**

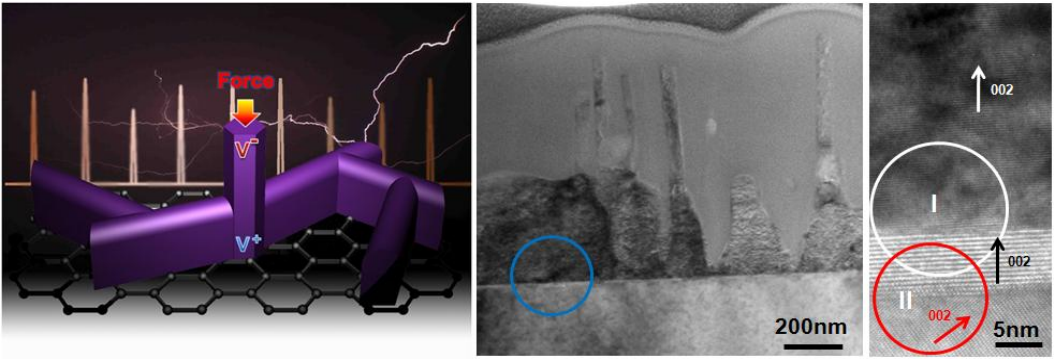

The morphologies of the ZnO nanostructures on graphene were controlled by varying the Au catalyst thickness and growth time via thermal chemical vapor deposition. Furthermore, we demonstrate that a DC output power can be generated by the piezoelectric ZnO/graphene nanogenerator due to the electron dynamic specific of the nanowire-nanowall hybrid structure.